\documentclass[showpacs,preprintnumbers,amsmath,amssymb]{revtex4}
\usepackage{bbold}
\usepackage{graphicx}
\usepackage{dcolumn}
\usepackage{bm}
\usepackage{slashed}

\begin{document}

\title{A non-associative quaternion scalar field theory}
\vspace{2cm}
\author{Sergio Giardino}
\email{giardino@ime.unicamp.br}
\affiliation{ Instituto de Matem\'{a}tica, Estat\'{i}stica e
Computa\c{c}\~{a}o Cient\'{i}fica, Universidade Estadual de Campinas\\
Rua S\'{e}rgio Buarque de Holanda 651, 13083-859, Campinas, SP,
Brazil}
\vspace{2cm}
\author{Paulo Teot\^{o}nio-Sobrinho}
 \email{teotonio@fma.if.usp.br}
\affiliation{Instituto de F\'{i}sica, Universidade de S\~{a}o Paulo,\\ CP 66318, 05315-970 S\~{a}o Paulo, SP, Brazil.}
\vspace{2cm}

\begin{abstract}
\noindent A non-associative Groenewold-Moyal plane is constructed
using quaternion-valued function algebras. The symmetrized
multi-particle states, the scalar product, the annihilation/creation algebra and the formulation in terms of a
Hopf algebra are also developed. Non-associative quantum algebras in
terms  of position and momentum operators are given as the  simplest examples
of a framework whose applications may involve string theory and non-linear quantum field theory. 
\end{abstract}

\maketitle

\section{Introduction}

Non-commutative geometry \cite{Connes:1994yd} has a wide range of
applications in quantum field theory
\cite{Akofor:2010jf,Szabo:2001kg}, in the construction of non-commutative physical models. 
These non-commutative theories are associative. A more general
framework could be conceived whereby, in addition to non-commutativity, the
algebra is also non-associative. Our aim is to find an example where 
non-commutative and non-associative algebra appears naturally in the
context of field theory. Since most field theories are based on associative
algebra, our aim is to obtain a deformation parameter 
$\theta$ such that associativity is recovered when $\theta$ goes to zero.

In the following pages, this goal is achieved by means of
construction:  we start with a field theory where the base space
is comprised of $\mathbb{R}^D$ and target space is comprised of quaternions $\mathbb{H}$. The
second step is to deform $\mathbb{R}^D$ into non-commutative algebra
such that $[x^\mu,x^\nu]=i\,\theta^{\nu\mu}$. It turns out that the
resulting algebra of fields is non-associative. As expected, when
$\theta^{\mu\nu}$ goes to zero, associativity is recovered.

Let us consider a  quaternion-valued field
theory, and write the field  $\mathcal{F}:\mathbb{R}^D\to\mathbb{H}$
in a symplectic notation as $\mathcal{F}=f_0+jf_1$, so that
$f_{i=0,1}:\mathbb{R}^D\to\mathbb{C}$. In this theory, the sources of 
non-commutativity are the quaternion complex units $i$, $j$ and
$k=ij=-ji$. By deforming the commutative multiplication of the
complex-valued functions $f_{i=0,1}$ to a non-commutative, we
obtain a theory with non-associativity as a byproduct of the
superposition of the these two different non-commutativities. 

Non-associative phenomena appear in many places, and further
information can be found in reviews on the subject
\cite{Okubo:2005n,Lohmus:1998ps}. However, while  non-associativity 
is common place in algebra 
\cite{Shafer:1996n}, examples of non-associativity in physics
are a collection of disconnected problems.
The most obvious proposals for finding a physical phenomenon that may be described by
non-associativity involve the octonion field
\cite{Kugo:1982bn,Foot:1988at,Gunaydin:1995as,Baez:2001dm,Gogberashvili:2005cp}. Although
octonion algebra is a standard example of non-associativity, it
does not have an  associative limit. 
 Recently, non-associative structures have appeared in general relativity,
\cite{Nesterov:1999mu,Nesterov:2000qb,Nesterov:2004nr,Sbitneva:2001es,Nesterov:2001pc},
string theory \cite{Blumenhagen:2010hj,Blumenhagen:2011ph,Mylonas:2012pg} and brane
theory 
\cite{Bagger:2006sk,Bagger:2007jr,Bagger:2008se,Gustavsson:2007vu,Yamazaki:2008gg,Jardino:2010it}. The
model proposed in this article is an attempt to obtain a very simple
example of non-associativity where associativity can be recovered at a
suitable limit.

The field theory described in this article has a natural interpretation since
its target space may be understood as the tangent space of a
hyper-complex manifold. In the same way a complex manifold is locally complex, a hyper-complex
manifold is locally quaternionic. In the context of super-symmetric
models, there can be various types of complex and hyper-complex
manifolds as found in super-symmetric extensions of non-linear sigma models
\cite{Zumino:1979et,AlvarezGaume:1980vs,AlvarezGaume:1980dk,AlvarezGaume:1981hm,Rocek:1983ja}, string compactification on $K3$ surfaces \cite{Aspinwall:1996mn}, generalized hyper-K\"{a}hler applied to string theory
\cite{Zabzine:2006uz,Ezhuthachan:2006yy}, and even  speculations on
the nature of time  \cite{Gibbons:2011jf}. Therefore, the model
presented here can be understood as a linearized version of such
non-linear sigma models.

Our results are related to quaternion quantum mechanics and quantum
field theory 
 \cite{Rotelli:1988fc,Gibbons:2011jf,Adler:1995qqm,Adler:1985uh,Adler:1985wz,DeLeo:1991mi,Gsponer:2005by}. However,
 these latter models do not consider multi-particle states, and consequently 
 in these theories it is impossible to build states with particle
 statistics, a problem that has been solved here by defining
 annihilation and creation operators of symmetrized states.

This paper is organized as follows: in the second section 
we present the non-deformed quaternion scalar
field theory and its multi-particle states and statistics. In the third
section a deformed  algebra of functions is formulated according to
 the Groenenwold-Moyal procedure. We then show that the resulting
 algebra is non-associative.  Examples of
non-associative quantum algebra obtained from introducing
quaternion unity are presented as well. The last section contains our
conclusions and future perspectives.
\section{The quaternion scalar field theory}
The afore mentioned quaternion field theories have only one-particle states.
This means that multi-particle states cannot be built according to
boson-fermion statistics. In this section this void is filled in
the mathematical structure of quaternion field theory following the
Hopf algebra formalism of \cite{Balachandran:2007vx}, where  
the Poincar\'{e} group acts on the Groenenwold-Moyal plane with a
deformed coproduct. In this section the deformation is the
hyper-complex quaternion structure. A second deformation, in the
usual multiplication, is introduced in the third section.

\subsection*{Poincar\'{e} invariance}

If $g$ is an element belonging to the Poincar\'{e} algebra, 
 the action of the symmetry algebra ($\triangleright$) on
space-time functions $\mathcal{F,\, G}\in\mathbb{H}$ must obey
\begin{equation}
g\triangleright(\mathcal{F\cdot G})=(g\triangleright\mathcal{F})\cdot(g\triangleright\mathcal{G}),
\end{equation}
 \noindent where the dot represents ordinary multiplication. It is
 adopted the symplectic notation for quaternionic functions, so that
 $\mathcal{F}=f_0+f_1j$, with $f_{i=1,2}$ $\mathbb{C}-$functions, and $j$ is the
 complex element of quaternion algebra, and thus $ij=-ji$. In
terms of Hopf algebras, the action of the elements of an algebra over a product of
complex functions is determined by the co-product. By way of example, the translation generator
$\hat{p}=i\partial_x$ of the Poincar\'{e} group acts on complex function
algebra according to the co-product 
\begin{equation}
\Delta(\hat{p})=\mathbb{1}\otimes \hat{p} + \hat{p}\otimes \mathbb{1},\label{coprod}
\end{equation}
\noindent which, acting on
$f,\,g\in\mathbb{C}$ with multiplication $m$, is subject to the consistency constraint
$m\big(\Delta(\hat p)(f\otimes g)\big)= \hat p(f\cdot g)$, where $m$
takes the elements of the tensor product and multiplies them. On the
other hand, taking the quaternion functions $fj$ and $gj$, again
with $f,\,g$ $\mathbb{C}-$functions, we obtain $m(\Delta(p)(f\otimes g))\neq p(f\cdot g)$. This difficulty is
solved by defining a quaternion tensor product, namely
\begin{eqnarray}
&(f \otimes g )\cdot(m \otimes n)=(f\cdot m)\otimes(g\cdot n),&\\
&(f \otimes gj )\cdot(m \otimes n)=(f\cdot \bar{m})\otimes(gj\cdot n),&\\
&(f \otimes g )\cdot(mj \otimes n)=(f\cdot mj)\otimes(\bar{g}\cdot n),&\\
&(f \otimes gj )\cdot(mj \otimes n)=(f\cdot \bar{m}j)\otimes(\bar{g}j\cdot n).
\end{eqnarray}
\noindent $f,\,g,\,m$ and $n$ are complex-valued functions and barred
functions are the complex conjugates. This result follows for
$f,\,n\in\mathbb{H}$ as well.
 This kind of structure is similar to the $\mathbb{Z}_2$
tensor product found in Lie super-algebras
\cite{Scheunert:1979sa}. Adopting this tensor product the 
co-product satisfies the identity
$\Delta(\hat{p}\,\hat{q})=\Delta(\hat{p})\Delta(\hat{q})$, and 
the first element of a multi-particle quaternionic state is given: a
well defined co-product. As the
co-product of the translation operator of the Poincar\'{e} algebra has
the expression (\ref{coprod}), either in the quaternion case or 
in the complex case, it will have the same behavior  when the
multiplication operation is deformed according to the Moyal procedure in
both cases. Thus the deformed co-product of the rotation
operator of the Poincar\'{e} group in the non-commutative complex
function algebra \cite{Chaichian:2004za} is valid for the
deformed quaternion spaces discussed in the next section as well.

\subsection*{State statistics}

States endowed with well-defined statistics have a permutation
operator which interchanges the positions of the functions
describing the particles in a state. As the particles are represented by
quaternion functions, for a generic quaternion state
\begin{equation}
\mathcal{F}\otimes\mathcal{G}=f_0\otimes g_0 + f_0\otimes g_1j +
f_1j\otimes g_0 + f_1j\otimes g_1j,
\end{equation}
\noindent the following operators are defined:
\begin{eqnarray}
&&\hat{\sigma}\triangleright(\mathcal{F}\otimes\mathcal{G})=(\mathcal{G}\otimes\mathcal{F})\\
&&\hat{\tau}\triangleright(\mathcal{F}\otimes\mathcal{G})=g_0\otimes
f_0 +g_1j\otimes \bar{f}_0 + \bar{g}_0\otimes f_1j + \bar{g}_1j\otimes \bar{f}_1j\\
&&\hat{\kappa}\triangleright(\mathcal{F}\otimes\mathcal{G})=f_0\otimes g_0 + \bar{f}_0\otimes g_1j +
f_1j\otimes \bar{g}_0 + \bar{f}_1j\otimes \bar{g}_1j,
\end{eqnarray}
\noindent so that $\hat{\sigma}^2=\hat{\tau}^2=\hat{\kappa}^2=\mathbb{1}$ and
$\hat\kappa=\hat\sigma\hat\tau$. Symmetrized states and
anti-symmetrized states are defined as follows
\begin{equation}
|\mathcal{F}_1,\,\mathcal{F}_2\rangle_{\pm} =\frac{1}{2}(\mathbb{1}\pm\hat{\tau})\triangleright|\mathcal{F}_1,\,\mathcal{F}_2\rangle,
\end{equation}
\noindent with $|\mathcal{F}_1,\,\mathcal{F}_2\rangle=
\mathcal{F}_1\otimes\mathcal{F}_2$, the defined states are
eigen-states of the permutation operator $\hat{\tau}$ according to
\begin{equation}
\hat{\tau}|\mathcal{F}_1,\,\mathcal{F}_2\rangle_{\pm} =\pm|\mathcal{F}_1,\,\mathcal{F}_2\rangle_{\pm}. 
\end{equation}
After defining symmetrized states and anti-symmetrized states, the
particle statistics is guaranteed, and a
scalar product is needed, which is presented below.
\subsection*{The scalar product}
By expressing a one-particle state as
$|\mathcal{F}\rangle=|f_0\rangle+|f_1j\rangle$, so that the
ortogonality condition $\langle f\,|\,gj\rangle= \langle
fj\,|\,g\rangle=0$ holds, a complex-valued scalar product is obtained
as a sum of usual 
scalar products of complex functions 
\begin{eqnarray}
\langle \mathcal{F}\,|\,\mathcal{G}\rangle&=&\langle f_0\,|\,g_0\rangle +\langle f_1j\,|\,g_1j\rangle\\
&=&\langle f_0\,|\,g_0\rangle +\langle g_1\,|\,f_1\rangle.\nonumber
\end{eqnarray}
\noindent In the above scalar product,
$\langle z\,\mathcal{F}\,|\,\mathcal{G}\rangle \neq \langle
\,\mathcal{F}\,|\,\bar{z}\,\mathcal{G}\rangle$, where $z\in \mathbb{C}$
and $\mathcal{F,\,G}:\mathbb{R}^4 \to \mathbb{H}$. As a consequence,
this fact will result in the splitting of the creation/annihilation operator
algebra, as shown in the next item. On the other hand,
when defining the scalar product of  two-particle states as
\begin{eqnarray}
\langle \mathcal{F},\mathcal{G}|\mathcal{M},\mathcal{N}\rangle
&=& \big(|\mathcal{F},\mathcal{G}\rangle, \,\hat{\kappa} |\mathcal{M},\mathcal{N}\rangle\big)\\
&=&\langle f_0|m_0\rangle \langle \mathcal{G}|\mathcal{N}\rangle +
\langle f_1 j|m_1 j\rangle \overline{\langle\mathcal{G}|\mathcal{N}\rangle},
\end{eqnarray}
\noindent it is observed that if $\mathcal{F}=\mathcal{M}$ and
$\mathcal{G}=\mathcal{N}$, then
\begin{equation}
|\mathcal{F}\otimes \mathcal{G}|^2=|\mathcal{F}|^2 | \mathcal{G}|^2.
\end{equation}
The scalar product also obeys the necessary self-adjointness condition
\begin{equation}
\big(\hat{\tau}|\mathcal{F},\mathcal{G}\rangle, \, |\mathcal{M},\mathcal{N}\rangle\big)=\big(|\mathcal{F},\mathcal{G}\rangle, \,\hat{\tau} |\mathcal{M},\mathcal{N}\rangle\big)\label{Tau}.
\end{equation}
\noindent Thus, the scalar product constructed above is valid for
multi-particles, something which has not been observed in previous quaternion quantum theories.
\subsection*{Creation and annihilation operators \label{aa+}}
In principle,  the creation $\mathbb{a}_\mathcal{F}^\dagger$ operator and
annihilation operator  $\mathbb{a}_\mathcal{F}$ of a quaternionic state are
\begin{equation}
\mathbb{a}_\mathcal{F}^\dagger=a_{f_0}^\dagger+a_{f_1j}^\dagger \qquad\mbox{and}\qquad\mathbb{a}_\mathcal{F}=a_{f_0}+a_{f_1j}.
\end{equation}
\noindent However, as the scalar product constructed above is such
that $\langle z\,\mathcal{F}\,|\,\mathcal{G}\rangle \neq \langle
\,\mathcal{F}\,|\,\bar{z}\,\mathcal{G}\rangle$, where $z\in \mathbb{C}$
and $\mathcal{F,\,G}:\mathbb{R}^4\to \mathbb{H}$, the creation/annihilation algebra will be built in
terms of  $a_{f_0}^\dagger$, $a_{f_1j}^\dagger $, $a_{f_0}$, and
$a_{f_1j}$. These operators create complex fields, and thus satisfy
commutation rules with the quaternionic unity $j$, namely 
\begin{equation}
za_{f_0}=a_{f_0}z,\qquad ja_{f_0}=a_{\bar{f}_0}j,\qquad
za_{f_1j}=a_{f_1j}\bar{z}\qquad\mbox{and}\qquad j a_{f_1j}=a_{\bar{f}_1j}j.
\end{equation}
\noindent The operator creates/annihilates either a bosonic or a
fermionic state, thus the wave-function must be either symmetrized or
anti-symmetrized. In order to construct the algebra, 
the scalar product must have the same result as that obtained by the creation
annihilation operators. The necessary scalar products are
\begin{eqnarray}
&&_\pm\langle f\otimes g,\,m\otimes n
\rangle_\pm=\langle f,\,m\rangle\langle g,\,n\rangle\pm\langle f,\,n\rangle\langle g,\,m)\\
&& _\pm\langle f\otimes gj,\,m\otimes nj
\rangle_\pm=\langle f,\,m\rangle\langle gj,\,nj\rangle\\
&&_\pm\langle fj\otimes gj,\,mj\otimes nj
\rangle_\pm=\langle fj,\,mj\rangle\langle\bar gj,\,\bar
nj\rangle\pm\langle fj,\,\bar n j\rangle\langle\bar gj,\,mj\rangle,
\end{eqnarray}
so that the plus sign corresponds to the symmetric bosonic states and
the minus sign corresponds to the fermionic anti-symmetric states. For
the bosonic case, the operator algebra reproduces the above results as
\begin{eqnarray}
&&[a_f,\,a_g]=[a_f^{\dagger},\,a_g^{\dagger}]=0\\
&&a_f\,a_{gj}- a_{gj}\,a_{\bar{f}}=a_f^\dagger\,a_{gj}^\dagger-a_{gj}^\dagger\,a_{\bar{f}}^\dagger=0\\
&&a_{fj}\,a_{gj}-a_{\bar gj}\,a_{\bar{f}j}=
a_{fj}^\dagger\,a_{gj}^\dagger-a_{\bar gj}^\dagger\,a_{\bar fj}^\dagger=0\\
&&a_{fj}a_{g}^\dagger- a_{\bar{g}}^\dagger  a_{fj}=0\\
&&[a_{f},\,a_{g}^\dagger]=\langle f|g\rangle\\
&&a_{fj}a_{gj}^\dagger- a_{\bar{g}j}^\dagger a_{\bar{f}j}=\langle fj|gj\rangle.
\end{eqnarray}
\noindent remembering that $f$ and $g$ are complex-valued
functions, and that $\langle a,\,zb\rangle=z\langle a,\,b\rangle$ and $\langle
a,\,zb\rangle=\bar z\langle a,\,jb\rangle$ are adopted. On the other hand, for an
anti-symmetric fermionic state, the operator algebra is
\begin{eqnarray}
&&\{a_f,\,a_g\}=\{a_f^{\dagger},\,a_g^{\dagger}\}=0\\
&&a_f\,a_{gj}+ a_{gj}\,a_{\bar{f}}=a_f^\dagger\,a_{gj}^\dagger+a_{gj}^\dagger\,a_{\bar{f}}^\dagger=0\\
&&a_{fj}\,a_{gj}+a_{\bar gj}\,a_{\bar{f}j}=
a_{fj}^\dagger\,a_{gj}^\dagger+a_{\bar gj}^\dagger\,a_{\bar fj}^\dagger=0\\
&&a_{fj}a_{g}^\dagger+ a_{\bar{g}}^\dagger  a_{fj}=0\\
&&\{a_{f},\,a_{g}^\dagger\}=\langle f|g\rangle\\
&&a_{fj}a_{gj}^\dagger+ a_{\bar{g}j}^\dagger a_{\bar{f}j}=\langle fj|gj\rangle.
\end{eqnarray}

\noindent  Thus the constructed quaternionic scalar field theory has all the
structure necessary:  Poincar\'{e} invariant one-particle and multi-particle
states, symmetrized and anti-symmetrized states with well-defined
statistics, a scalar product and a creation/annihilation operator
algebra. This theory can be deformed according to the Groenenwold-Moyal
procedure generalizing the well-known non-commutative
complex field theories, and this is carried out in the next section.

\section{The deformed product}
Non-commutative geometry is obtained by changing the ordinary
commutative product of complex-valued functions  $f$ and $g$ into the
Groenenwod-Moyal (GM) deformed product
\begin{equation}
f(x)\star g(x)=f(x)g(x)+\sum_{n=1}^{\infty}\left(\frac{i}{2}\right)\frac{1}{n}\,\theta^{i_1j_1}\dots\theta^{j_nj_n}\,\partial_{i_1}\dots\partial_{i_n}f(x)\,\partial_{j_1}\dots\partial_{j_n}g(x),
\end{equation}
\noindent so that $\theta^{ij}$ is anti-symmetric in its
indices. Linear functions generate the commutator between coordinates
\begin{equation}
x^{i}\star x^{j} - x^{j}\star x^{i} = \theta^{ij},
\end{equation}
\noindent and in the limit where $\theta^{ij}\to 0$ the commutative
geometry is recovered. Both the commutative product and the
non-commutative product of functions are associative.

A more general picture may be obtained by deforming the usual product
of quaternion-valued functions according to the GM prescription.
The quaternionic-valued functions $\mathcal{F}:\mathbb{R}^D\to\mathbb{H}$ over a
$D-$dimensional Euclidean space with coordinates $x^i$ are
represented by
\begin{equation}
\mathcal{F}=f_1+f_2\,j,\qquad\mbox{so that}\qquad f_{a=1,2}:\mathbb{R}^D\to\mathbb{C}.
\end{equation}
\noindent $f_a$ are defined on a Schwarz space, thus allowing a
Fourier transform $\tilde{\mathcal{F}}$, where
\begin{equation}
\tilde{\mathcal{F}}=\tilde{f}_1+\tilde{f}_2\,j,\qquad\mbox{and}\qquad
\tilde{f}_a(k)=\int d^Dx\,e^{-ik_\mu x^\mu}f_a(x).
\end{equation}
\noindent Accordingly, the Weyl symbol of a quaternion function may
be introduced as well, so that
\begin{equation}
\hat{W}[\mathcal{F}]=\hat{W}[f_1]+\hat{W}[f_2]j,\qquad\mbox{and}\qquad
\hat{W}[f_a]=\int \frac{d^Dk}{(2\pi)^D}e^{-ik_\mu x^\mu}\tilde{f_a}.
\end{equation}
\noindent The Weyl symbol allows the GM product to be introduced, thus replacing the usual multiplication, so that the complex-valued functions
obey $\hat{W}[f_a\star f_b]=\hat{W}[f_a]\star \hat{W}[f_b]$, which
results in
\begin{equation}
\mathcal{F}\star \mathcal{G}=f_1\star g_1 +(f_1\star
g_2)j+j(\bar{f}_2\star g_1)+j\big((\bar{f}_2\star g_2)j\big),\label{FsG}
\end{equation}
\noindent where the bar means the conjugate of the complex
function. This star product of quaternion functions is, of course,
non-commutative; nevertheless, it is also non associative, so that
\begin{equation}
(\mathcal{F}\star \mathcal{G})\star \mathcal{H} - 
\mathcal{F}\star (\mathcal{G}\star \mathcal{H}) \neq 0.\label{FGH}
\end{equation} 
This is an interesting byproduct for introducing a non-commutative
local structure on a former non-commutative complex structure. This
simple theory has a number of possible applications, as cited in
the introduction of this paper.
\subsection*{Non-associative quaternion quantum algebras}
The simplest example of a non-associative deformed theory comes from
quantum mechanics and its celebrated commutation relation
\begin{equation} [\hat{x},\hat{p}]=i\hbar,
\end{equation}
whose $\hbar\to 0$ limit, or classical limit, turns the operators into a
commutative algebra. Introducing the quaternion complex unity $j$ naturally generates a
non-associative structure. As $j$ does not commute with $ [\hat{x},\hat{p}]$,
it does not associate with the products of the commutator anymore. The associator
$(\hat{x},\,j,\,\hat{p})=(\hat{x}\,j)\,\hat{p}-\hat{x}\,(j\,\hat{p})$
may be calculated in  the specific case where the quantum quaternion
algebra is an alternative algebra. Using the
Moufang identities \cite{Shafer:1996n}, the resulting associator is
\begin{equation}
(\hat{x},\,j,\,\hat{p})=k\hbar.\label{xjp}
\end{equation}
\noindent so that $k=ij$. This example in which quantum
mechanics turns to a non-associative theory is somewhat surprising,
but it shows very simply how combining non-commutative structures
generates a non-associative one. In this case, the associative limit
goes to a commutative complex theory, but this is a classical one. In
this sense, the commutativity and associativity are coupled. A
non-coupled case comes in the more general framework discussed previously.

On the other hand, it is possible to
further extend the quantum algebra. Defining the operators
\begin{equation}
\hat z^\dagger=\frac{1}{\sqrt{2}}\big(p+ix\big)\qquad\mbox{and}\qquad \hat z=\frac{1}{\sqrt{2}}\big(p-ix\big)
\end{equation}
so that $[\hat z,\,\hat z^\dagger]=\hbar$, and with the use of the
associator (\ref{xjp}), 
\begin{eqnarray}
\big(\hat z,\,j,\,\hat z^\dagger\big)=(\hat z^\dagger,\,j,\,\hat
z\big)=\big(\hat z,\,\hat z,\,j\big)=\big(j,\,\hat z,\,\hat z\big)=\big(\hat z^\dagger,\,\hat z^\dagger,\,j\big)=\big(j,\,\hat z^\dagger,\,\hat z^\dagger\big)=0&&\\
\big(\hat z^\dagger,\,\hat z,\,j\big)=-\big(j,\,\hat z^\dagger,\,\hat z\big)=-\big(\hat z,\,\hat z^\dagger,\,j\big)=\big(j,\,\hat z,\,\hat z^\dagger\big)=-\big(\hat z,\,j,\,\hat z\big)=\big(\hat z^\dagger,\,j,\,\hat z^\dagger\big)=j.&&
\end{eqnarray}
This is also a non-associative and non-commutative algebra, although
it is not alternative as that formed by  $\hat x$, $\hat p$ and
$j$, but its classical limit is also a quaternion classical quaternion
theory as expected.
The above examples are the simplest examples of the deformed
algebras, whose geometry is to be analyzed in forthcoming studies.
\section{Conclusion}

In this article two novel quaternion quantum scalar field theories have
been presented. Both of them are non-commutative because of the
quaternion nature of their fields. In one of them ordinary
commutative multiplication is defined, and in this case a
multi-particle quaternion scalar theory has been constructed. The second theory is a
deformation of the former one according to the Groenenwold-Moyal
procedure. This second theory is a non-commutative and non-associative
 one. These theories are well-defined, and may be used in a number of
physical applications, as the models are quite general. Developments in 
quaternion scalar fields and non-associative geometry are the
most immediate applications. We expect that results derived from this
linear model  will be useful when applied to hyper-K\"{a}hler structures
in string theory and non-linear sigma models.  

{\bf Acknowledgements} Sergio Giardino is grateful for the support
offered by the Departamento de F\'{i}sica Matem\'{a}tica of the Universidade de
S\~{a}o Pulo and also for the financial support provided by
Capes. Both of the authors  thank B. Chandrasekhar, 
A. P. Balachandran and T. R. Govindarajan for invaluable discussions.

%
%
%
%
\bibliographystyle{unsrt} 

%
%
%
%
\end{document}